\documentclass{ws-ijmpd}
\newcommand{\met}{\hbox{E\kern-0.5em\lower-0.1ex\hbox{/}}_T}
\newcommand\simlt{\lower.5ex\hbox{$\; \buildrel < \over \sim \;$}}
\newcommand\simgt{\lower.5ex\hbox{$\; \buildrel > \over \sim \;$}}

\begin{document}

\markboth{Amir Levinson}
{Instructions for Typing Manuscripts (Paper's Title)}

%
\catchline{}{}{}{}{}
%

\title{JETS ON ALL SCALES}

\author{Amir Levinson}

\address{Raymond and Beverly Sackler School of Physics and Astronomy, Tel Aviv University\\
Tel Aviv 69978, Israel\\
Levinson@wise.tau.ac.il}

\maketitle

\begin{history}
\received{Day Month Year}
\revised{Day Month Year}
\comby{Managing Editor}
\end{history}
    
\begin{abstract}
A brief overview of jets and their central drivers is presented, with a focus on accreting black hole systems.
In particular, scaling relations that elucidate some basic properties of the engines are derived, and the
implications for the associated outflows are discussed.  The kinematics and dynamics of relativistic jets in various systems and 
the dissipation of their bulk energy is considered, with an emphasis on consequences of recent observations.
Also considered is the interaction of the jets with their environment.  Comments on multi-messenger probes
are made at the end.
\end{abstract}

\keywords{Jets; blazars; microquasars; gamma-ray bursts}
\section{Introduction}
Advances of observational techniques and improvements in numerical capabilities in the last decade or so have 
led to a progress in our understanding of astrophysical jets and their drivers, 
but also raised new and highlighted some old questions, mainly with regards to the micro-physics involved.
For instance, the physics of accretion, the details of the Blandford-Znajek mechanism, the acceleration 
of ideal MHD outflows, the physics of collisionless shocks and the role of relativistic radiation mediated shocks 
are better understood now then a decade ago.  On the other hand, the loading and dissipation of magnetically 
dominated flows, the generation of magnetic fields behind shocks, as implied by observations of afterglow emission in GRBs, the role of relativistic 
turbulence in and its effects on the dynamics of the expanding flow, and the nature of the compact object in certain systems 
are but some examples of open questions that only very recently have started to be examined systematically using advanced tools.
We are still lacking any knowledge of the composition of relativistic jets in almost all sources, do not 
understand yet how magnetic flux is advected from large radii all the way into the very inner regions of the disk
in AGNs and microquasars (or is it produced locally by some mechanism? e.g., a Poynting-Robertson battery\cite{Coetal09}), 
are puzzled by the detection 
of UHECRs, and hope for detection of VHE neutrinos from the jets and gravitational waves from their engines that will 
shed a new light on some of these open questions.

Below is a brief summary of some of these issues, that reflects my naive perception of the field.

\section{Central Engine}	
A common view is that the launching of astrophysical jets involves a rotating, magnetized driver.  In compact relativistic 
systems the central engine may consist of a magnetized neutron star, as in the case of pulsars, $\gamma$-ray binaries, 
and magnetars, or an accreting black hole, as in blazars, microquasars and some classes of GRBs.  The following discussion
focuses on accreting black hole systems. 
\subsection{Scaling of conditions in the disk}
Two important parameters control the conditions in the inner regions of a disk surrounding an accreting black hole: the 
mass of the black hole, henceforth measured in units of solar mass, $m_{BH}=M_{BH}/M_\odot$, and the accretion rate
$\dot{m}$, rendered dimensionless by measuring in Eddington units $\dot{M}_{Edd}=L_{Edd}/c^2$.  The temperature, density 
and magnetic field strength scale with the ratio $\dot{m}/m_{BH}$ as
\begin{eqnarray}
T_d&=&10^7(\dot{m}/m_{BH})^{1/4}(r/3r_s)^{-3/4}\qquad{\rm K},\\
\rho_d&=&10^{-6}\alpha^{-1}(\dot{m}/m_{BH})(r/3r_s)^{-3/2}\qquad{\rm gr\ cm^{-3}},\\
B&=&10^{8.5}(\xi_B\dot{m}/m_{BH})^{1/2}\qquad{\rm G}.\label{Bd}
\end{eqnarray}
The two additional parameters that appear in the above scaling, the viscosity parameter $\alpha$ and the 
magnetization $\xi_B$, represent parametrization of poorly understood micro-physics.  Numerical simulations 
seem to indicate that their values are essentially independent of $\dot{m}$ and $m_{BH}$ and span a rather 
narrow range.

The choices of canonical values $\dot{m}=1$, $m_{BH}=3$ and $\dot{m}=1$, $m_{BH}=10^8$, representing a
prototypical microquasar and a prototypical blazar, respectively, yield disk temperatures that are consistent with the peak of the SED.
For GRBs, with $\dot{m}=10^{15}$, $m_{BH}=3$, a disk temperature of a few MeV and a density exceeding
$10^{10}$ gr cm$^{-3}$ is anticipated.  In this regime the weak interaction time scale becomes comparable to
the accretion time and the inner regions of the disk cools via emission of MeV neutrinos,
and may contain neutron rich material \cite{chen03}.  A neutron-to-proton ratio in excess of 20 can be reached 
under certain conditions in the innermost regions.  If picked up by the GRB producing jet, such a neutron rich composition may have important 
consequences for the loading of the flow \cite{LE00} and for the prompt emission mechanism 
\cite{LE00}\cdash\cite{bel09}.  Whether the outflow can 
remain neutron rich as it accelerates to high Lorentz factors is yet an open issue \cite{BL08,MTQ08}.  
It could well be that the flow is multi-component, consisting of an ultra-relativistic core ensheathed by 
a slow neutron rich wind.  Leaking of free neutrons (that can easily cross magnetic field lines) 
from the slow wind into the baryon poor core can initiate a nuclear avalanche
that leads to baryon loading of the inner flow and dissipation of its bulk energy.  A hard gamma-ray and neutrino spectrum
is then expected\cite{LE00}.

\subsection{The disk-outflow connection}
MHD simulations seem to indicate that magnetic launching of a relativistic outflow in accreting black hole systems requires 
the presence of a Kerr BH with a specific angular momentum $\tilde{a}$ not much smaller than unity.  Turbulence in the disk
leads to a rapid redistribution of magnetic field lines and substantial mass loading, and so outflows from the disk are expected to be
slow, unlike the Blandford-Pyne solutions.
                   
The power that can, in principle, be extracted magnetically from a rotating black hole can be expressed as 
\begin{equation}
L_{j}=10^{21}\epsilon \tilde{a}^2B^2m_{BH}^2 \qquad {\rm erg\ s^{-1}},\label{LBZ} 
\end{equation}
where  $\tilde{a}$ is the specific angular momentum of the hole, and $\epsilon$ is a parameter that depends
on magnetic field geometry and other details.  By employing (\ref{Bd}) a simple relation between the accretion rate 
and outflow power is obtained:
\begin{equation}
L_{j}=(\xi_B\epsilon\dot{m})L_{Edd}.\label{LBZ2} 
\end{equation}
This scaling appears to be consistent with the outflow power inferred in different classes of sources.  However,
the correlation between outflow ejection and spectral states in X-ray binaries reveals high accretion states during
which the outflow is strongly suppressed, suggesting that additional effects may be involved. 

The disk luminosity has a similar scaling,
\begin{equation}
L_{d}=(\xi_r\dot{m})L_{Edd},\label{Ldr} 
\end{equation}
here $\xi_r$ is the radiative efficiency of the accretion flow. The presence of powerful $\gamma$-ray flares in some 
blazars suggests that in some circumstances the accretion mode is radiative inefficient.  An example is 
the extreme flare reported for PKS 2155-304.  The flare duration, $t_{\rm var}=300$ sec, and the isotropic 
equivalent luminosity, $L_{TeV}\simgt10^{46}$ erg s$^{-1}$, imply 
\begin{equation}
L_j>f_bL_{TeV}\simeq10^{44}\theta_{-1}^2L_{TeV,46} \qquad {\rm erg\ s^{-1}},\label{L_j} 
\end{equation}
with $f_b=\theta^2/2$ denoting the beaming factor of the emission for a two-sided conical jet with an opening 
angle $\theta=0.1\theta_{-1}$.  To avoid $\gamma\gamma$ absorption of the observed TeV photons by the disk 
radiation at small radii, that would smear out any rapid variations, requires either unusual beaming, or 
low radiative efficiency $\xi_r\simlt10^{-3}$ for typical opening angles.   

Perhaps the best example of radiative inefficient accretion is M87.
Various estimates of the jet power yield $L_j\simgt10^{44}$ erg/s (see Ref.~\refcite{Lietal09} for a summary of 
published estimates), implying $\dot{m}\sim10^{-2}$.  The bolometric
luminosity, on the other hand, is smaller by a factor of $10^{-3}$, suggesting $\xi_r\simlt10^{-3}$.  
It is worth-noting that the luminosity emitted from the jet itself is a small fraction of the jet power,
so this object is in fact a good example of a ``dark'' source.  This seems to be quite common 
among BL Lac sources.  Whether powerful, dark blazars are present in the nearby Universe is a question of interest  
in connection with potential UHECRs sources, as explained below.

\section{Jets}
\subsection{Kinematics and dynamics}
The best indication that jets associated with compact astrophysical systems are relativistic is of course 
the measurement of superluminal motions, which reflect the speed of some pattern, not necessarily the fluid,
that propagates down the jet.  The range of values inferred for the associated Lorentz factors
is $\Gamma\sim1-50$ in blazars and $\Gamma\sim1-10$ in microquasars.

Constraints on the Lorentz factor of emitting fluid are commonly derived using opacity arguments.  Those are 
mainly applied to GRBs and blazars in which the contribution of ambient radiation can be neglected
on relevant scales.   If both the gamma-rays and the target photons are produced isotropically inside a
source moving at a Lorentz factor $\Gamma$, e.g., by synchrotron and SSC mechanisms, 
then both components will be beamed into a cone of opening semi 
angle $\theta\sim\Gamma^{-1}$ in the star frame.  Two factors then lead to suppression of the pair production
opacity:  firstly, the flux factor, as measured in the star frame, satisfies $(1-\cos\theta)\sim1/(2\Gamma^{2})$.  Secondly, the threshold
condition implies that only target photons having energy $\epsilon_s>4\Gamma^2/\epsilon_{\gamma}$ can absorb
a $\gamma$-ray photon of energy $\epsilon_{\gamma}$ (energies are measured in units of $m_ec^2)$, 
and so the number density of target photons above the 
threshold is $n_s(4\Gamma^2/\epsilon_{\gamma})\propto \Gamma^{-2\alpha}$ for a power law energy distribution 
of the form $n_s(\epsilon_s)=K(r)\epsilon_s^{-\alpha}$.  The pair production optical depth then scales as 
$\tau_{\gamma\gamma}\propto\Gamma^{-2(\alpha+1)}$.
Now, the size of the emission zone $r_{em}$ may be constrained by variability of the observed flux.  For a relativistic 
source $r_{em}\le\Gamma^2 \Delta t /c$, where $\Delta t$ is the shortest variability 
time observed at energy $\epsilon_\gamma$.  The requirement that $\tau_{\gamma\gamma}<1$ at $r=r_{em}$ 
then yields, assuming $K(r)=K_0(r/r_0)^{-2}$,
\begin{equation}
\Gamma>\Gamma_{min}=\left(\frac{3\pi K_0A(\alpha)}{8\Delta t}\right)^{1/(2\alpha+4)}{\epsilon_\gamma}^{\alpha/(2\alpha+4)}
\end{equation}
where $A(\alpha)$ is a numerical factor that depends on the exponent $\alpha$, and is given in Ref.~\refcite{BL95}.  
For typical values of $\alpha$ $A(\alpha)$ lies in the range 0.1-0.2.
The observables $\epsilon_\gamma, \Delta t$ and the observed luminosity that fixes $K_0$ impose a constraint
on $\Gamma$.  Such opacity arguments have been applied to GRBs, whereby $\Gamma\sim10^3$ \cite{Gretal09} has been inferred in
the most extreme cases, and to TeV blazars \cite{Le06}, where $\Gamma>50$ have been estimated for several sources.

What mechanism accelerates the flow to such high Lorentz factors? Magnetic acceleration is one possibility. 
In general it is not very effective in the sense that the flow remains asymptotically Poynting dominated.  For
a split monopole $\Gamma_{\infty}\simeq\sigma^{1/3}$ where $\sigma$ is the ratio of magnetic to kinetic energy
at the base of the flow.  However, it has been shown recently that causal sections can be magnetically accelerated 
up to equipartition where $\Gamma_\infty\sim\sigma$ \cite{Kometal09,Lyu09}.  In case of GRBs the opening angle naively anticipated for 
the asymptotic flow, $\theta\simlt\Gamma_\infty^{-1}$, seems to be significantly smaller than those inferred from 
observations.  The latter condition may be alleviated in outflows that break out of a star\cite{TNM09}, as anticipated in long GRBs. 
We note that $\Gamma\sim10^3$ has been reported recently for some short bursts (e.g., GRB090510).

In outflows having a large Thomson depth the radiation is strongly coupled to the plasma.  If the entropy per baryon
at the base of the flow is large then the flow can radiatively accelerate to a large terminal Lorentz factor $\Gamma_{\infty}$
that depends on the location of the photosphere with respect to the coasting radius.  For a burst of total energy $E$ and
baryon mass $M_b$ the terminal Lorentz factor is $\Gamma_{\infty}\simeq E/M_bc^2$ if the photospheric radius 
$r_{ph}$ is larger than the coasting radius  $r_c$.  In the opposite limit, $r_{ph}<r_c$, the terminal Lorentz factor roughly 
satisfies $\Gamma_\infty\simeq\Gamma_0(r_{ph}/R)$, where $\Gamma_0\sim1$ is the Lorentz factor at the base of the 
outflow, at $r=R=10^6 R_6$ cm.  At the critical loading for which $r_{ph}=r_c$ the asymptotic Lorentz factor is given by
\begin{equation}
\Gamma_c\simeq1.8\times10^3 L_{52}^{1/4}R_6^{-1/4},
\end{equation}
with $L_{52}$ being the isotropic equivalent luminosity in units of $10^{52}$ erg s$^{-1}$.
Detection of sources that violate this limit would strongly support magnetic acceleration. 
For GRB 080916C we estimate $\Gamma_c\sim5500$.  

\subsection{Dissipation}
Dissipation of the outflow bulk energy occurs over a large range of scales.  It can be accomplished trough
overtaking collisions of fluid shells (internal shocks), as a result of interactions of the outflow
with a surrounding medium (e.g., recollimation shocks, breakout shocks, blast waves) or, in magnetically 
dominated regions, due to magnetic reconnection and/or instabilities.  

Shocks that form by overtaking 
collisions can dissipate energy at radii $r_d>\Gamma^2c\delta t$, where $\Gamma$ is the Lorentz factor of the slow shell
and $\delta t\ge r_s/c$ is the duty cycle of the intermittent engine.  In blazars and microquasars with $\Gamma\sim 1-50$ 
dissipation by internal shocks is expected close to the BH, consistent with (but not necessarily implied by) the 
short durations of strong flares observed in these objects, particularly in TeV blazars.  

In GRBs $r_d>10^5 r_s$ 
or so for the Lorentz factors envisaged.  For the shocks to form above the photosphere 
$\Gamma>200 L_{52}^{1/5}\delta t^{-1/5}_{-3}$ is required, where $\delta t_{-3}=\delta t/(1 {\rm ms})$.  In case of
GRB 080916C, for which $L_{52}\sim 100$ was measured during the first few seconds, this implies $\Gamma>800$ 
in order that the prompt emission be produced in optically thin regions. This value is
comparable to the limit derived using opacity constraints on the highest GeV photons recorded, as explained above.
Thus, it seems that in this burst a sizable fraction of the available energy may dissipate slight above or 
just below the photosphere, in regions 
of modest Thomson depth, $\tau\sim 1 -10^2$.  Shocks that form above the photosphere are expected to be collisionless.  These can 
Fermi accelerate particles and produce nonthermal spectra with a modest efficiency. 
Shocks that form below the photosphere, where the Thomson depth exceeds unity, 
are mediated by Compton scattering \cite{BoL08,KBW09}.   Under conditions anticipated in GRBs these shocks 
convect enough radiation upstream to render photon production in the shock transition negligible\cite{BL209}.  
Bulk Comptonization then produces a broad, nonthermal component
in the immediate downstream that extends up to a fraction of the KN limit in the shock frame, depending on
details (or up to a fraction of $\sim \Gamma m_e c^2$ in the observer frame).   At what depth thermalization  is established 
is yet an open issue.  We naively expect the spectrum to be
quasi thermal if the Lorentz factor is sufficiently small to allow shocks to form well below the Thomson sphere, and nonthermal if
a considerable fraction of the energy dissipates in a region where the Thomson depth is modest (less than a few hundreds). 
According to this interpretation the lack of a thermal component in the 
prompt emission from GRB 080916C implies that shocks are produced by radiation mediated shocks that  
form at a modest Thomson depth, consistent with the limit derived on the Lorentz factor.   Alternatively, the shocks
are collisionless. 
The recent detections of some bursts that exhibit a prominent thermal component suggests that in those sources dissipation occurred
deep enough below the photosphere, on thermalization scales.   The nonthermal extension requires additional dissipation 
above or just below the photosphere.


\subsection{Interaction with the Environment}
The environment plays an important role both through direct interactions with the jet and/or
by screening the jet emissions.  Collimation and blast waves/cocoons are generic environmental 
signatures in all sources.  This interaction may provide an important heating mechanism of IGM gas in clusters.

In microquasars associated with a massive companion the hydrodynamic and 
emission from the jet may be dominated by interactions with the wind and radiation from the stellar companion
(for a review see Ref.~ \refcite{BRK09} and references therein). 
Even the nature of the compact object in at least two TeV microquasars, LS 5039 and LS I+61 303, is controversial\cite{BRK09}. The recent 
detection of GeV emission \cite{Fermi1,Fermi2} from these two sources clearly indicates two components, one that peaks at a few GeV
and a second one extending to TeV energies.  Both components show modulations consistent with the orbital motion of 
the binary system (with a phase difference between the peak flux of each of the components), indicative of the interaction with the companion star.
In both objects the modulation of the GeV emission appears to be consistent with IC scattering of the companion's radiation; the suppression
of the TeV flux during the peak of the GeV emission may be due to enhanced pair production opacity. 
Alternatively, the GeV emission may originate from a pulsar magnetosphere\cite{Fermi2}, however, in this case the modulation of
the flux requires additional explanation.

In long duration GRBS the jet interacts with the putative stellar envelope.  A successful event requires breakout of the jet
from the star.  Owing to the scaling of the velocity at which the jet head advances with the expelled power, a successful breakout 
favors low power jets (for a given explosion energy), so that it could well be that in case of GRBs associated 
with collapsars long events are pre-selected by the environment.  Failed GRBs may have a different appearance.  In 
particular, an orphan burst of VHE neutrinos may be a unique diagnostic of chocked outflows\cite{MW01}, provided the Lorentz 
factor of the hidden jet is sufficiently high to render internal shocks that form in the jet collinsionless\cite{BoL08}, 
which is required for efficient acceleration of the protons that interact with the radiation produced behind the bow 
shock.

The subsequent interaction of the jet with a stellar wind or ISM produces a relativistic blast wave.  
The post-prompt emissions observed in most long GRBs are most likely produced 
in the thin layers enclosed between the forward and reverse shocks, and are important diagnostics
of the blast wave evolution and the conditions in the shocked layers.  Although a simple blast wave
model has been quite successful in explaining the late afterglow evolution,
recent observations raise some questions. In particular, 
(i) observations of the late afterglow emission 
indicate strong amplification of magnetic fields in the post shock region - by several orders of
magnitudes larger than what can be achieved by compression of the ambient magnetic field. Kinetic instabilities
have been proposed as the origin of these magnetic fields, however, whether the resulting fields can be maintained
over sufficiently large scales is yet an open issue.  An alternative is amplification by turbulence. 
(ii) SWIFT observations during the early afterglow 
phase reveal strong deviation of the lightcurve at early times from that predicted by the simple blast wave model. 
Several {\em ad hoc} explanations have been offered, including prolonged activity of the central engine 
and evolution of microphysical parameters.  However, the feasibility of these scenarios depends on poorly understood physics, and
it remains to be demonstrated that they can be derived from first principles.
(iii)  In the fireball scenario commonly adopted,
the naive expectation has been that the crossing of the reverse shock should produce an observable 
optical flash.  Despite considerable observational efforts, such flashes seem to be very rare. It could be
that the ejecta is magnetically dominated \cite{LE93,GMA08}, though it is not clear at present how a thin magnetic
shell can reach such large radii without expanding considerably.  Moreover, 
some accumulation of baryon rich matter at the 'piston's' head is anticipated during the shock 
breakout phase, that may mimic effects of a hydrodynamic ejecta.

Recently\cite{Le09a} it has been shown that the contact discontinuity of the decelerating shell is
unstable to convective Rayleigh-Taylor modes having angular scales smaller than the causality scale. 
It has been speculated that the convective instability may be an inherent source of turbulence in
the shocked circumburst layer that leads to a strong amplification of magnetic fields over a long portion of 
the blast wave evolution. 
The linear stability analysis also indicates a rapid response of the reverse shock to distortions at the contact, 
suggesting that the instability can affect the emission 
from the shocked ejecta in the early post-prompt phase of GRBs, and may be the reason for the apparent
lack of optical flashes.

\section{Multi-messenger probes}
Multi-messenger emissions carry important information that is not accessible to electromagnetic radiation,
primarily because the innermost regions of the relativistic outflows and their engines 
are opaque to electromagnetic radiation.  Recent and future experiments, specifically LIGO and EGO, 
cubic km neutrino telescopes, and the Auger UHECR experiment will hopefully advance our understanding 
further.  Detection of gravitational waves from GRBs for instance can be used to probe the innermost
region of accreting Kerr holes \cite{Vanetal04}.  Detection of VHE neutrinos will pin down the composition of the jets, and
will provide more stringent constraints on particle acceleration.  Optimistic estimates suggest that 
blazars\cite{AD01}, microquasars\cite{LW01,Dsetal02,Chetal06} and GRBs\cite{WB97} may all be detectable
by cubic km neutrino telescopes under optimal conditions.  The association of UHECRs with any astrophysical 
source has already interesting implications, as discussed below.

The origin of UHECRs (those above the ankle) is still a mystery.  It is widely believed that the
sources are extragalactic, though they have not yet been identified.
The confirmation of a GZK feature in the data strongly supports a bottom-up scenario, as otherwise 
such a scale would appear as a peculiar coincidence.   There is some evidence for a weak anisotropy in the
arrival directions of UHECRs events \cite{Auger08} that suggests a correlation of
the UHECRs sources with the large-scale structure in the local Universe \cite{Kash08}.
A general constraint on UHECRs sources can be derived from the requirement that the accelerated particles
are confined to the acceleration region; specifically that the escape time $t_{\rm esc}=r/c\Gamma$ is longer
than the acceleration time $t_{\rm acc}\simeq r_{L}(\epsilon)/c$, where $r_L(\epsilon)$ is the Larmor radius of
a particle having energy $\epsilon$.  This gives a relation between the source size and the strength of magnetic
field that depends to some extent on the composition of UHECRs.  Under the assumption that the 
UHECRs are accelerated in a relativistic magnetized outflow this also implies a minimum outflow power
\begin{equation}
L_j>10^{46} \Gamma^2\left(\frac{\epsilon}{10^{20} {\rm eV}}\right)^2 \qquad {\rm erg\ s^{-1}}.   
\label{Lj}
\end{equation} 
From (\ref{LBZ2}) it is seen that AGNs with $m_{BH}\simgt 10^9$, $\dot{m}\sim1$ and GRBs can account for 
the required power.  Strongly magnetized ($B> 10^{14}$ G) neutron stars are also potential candidates.  
The lack of bright AGNs within the GZK sphere implies the existence of dark blazars if the UHECRs indeed
originate from such objects.  A total radiative efficiency of the order of that inferred in M87 is  
sufficiently small to satisfy observational constraints. 

The condition (\ref{Lj}) should not necessarily apply in cases where the UHECRs are accelerated in regions 
that violate ideal MHD, e.g., starved black hole magnetospheres in dormant AGNs \cite{BG99,lev00} or 
boundary shear layers in subrelativistic jets\cite{RA09}.  The former scenario predicts a deletable, 
magnetospheric TeV emission owing to curvature losses\cite{lev00}.  With the new generation ICTA it should
be possible to test this hypothesis with a high statistical significance.

\section*{Acknowledgments}
Support by an ISF grant for the Israeli Center for High Energy Astrophysics is acknowledged.


\begin{thebibliography}{0}    
\bibitem{Coetal09} I. Contopoulus, et al. {\it Astrophys. J.}, {\bf 702} (2009) L148
\bibitem{chen03} W-X. Chen and A.~M.  Beloborodov, {\it Astrophys. J.}, {\bf 657} (2007) 383
\bibitem{LE00} A. Levinson and  D. Eichler, {\it Astrophys. J.}, {\bf 594} (2003) L19
\bibitem{BM00} J.~N. Bahcall and P. Meszaros, {\it Phys. Rev. Lett.}, {\bf 85} (2000) 1362
\bibitem{bel09} A.~M.  Beloborodov,  {\it arXiv:0907.0732} (2009)
\bibitem{BL08} Y. Barzilay and A. Levinson, {\it New Astron.}, {\bf 13} (2008) 386
\bibitem{MTQ08} B.~D. Metzger, T.~A. Thompson and E. Quataert, {\it Astrophys. J.}, {\bf 676} (2008) 1130
\bibitem{Lietal09} L. Yan-Rong, et al. {\it Astrophys. J.}, {\bf 699} (2009) 513

\bibitem{BL95} R.~D. Blandford and A. Levinson, {\it Astrophys. J.}, {\bf 441} (1995) 79
\bibitem{Gretal09} J. Granot, et al. {\it arXiv:0905.2206} (2009) 
\bibitem{Le06} A. Levinson, {\it Int. J. Mod. Phys. A}, {\bf 21} (2006) 6015
\bibitem{Kometal09} S. Komissarov, et al. {\it Mon. Not. Roy. Astron. Soc.}, {\bf 394} (2009) 1182
\bibitem{Lyu09} Y. Lyubarsky, {\it arXiv:0909.4819} (2009)
\bibitem{TNM09} A. Tchekhovskoy, R. Narayan and J.~C.  McKinney,  {\it arXiv:0909.0011} (2009)


\bibitem{BoL08}  A. Levinson and O. Bromberg, {\it Phys. Rev. Lett.}, {\bf 100} (2008) 131101
\bibitem{KBW09} B. Katz, R. Budnik and E. Waxman {\it arXiv:0902.4708} (2009)
\bibitem{BL209}  O. Bromberg and A. Levinson, in preparation

\bibitem{BRK09} V. Bosch-Ramon and D. Khangulyan, {\it Int. J. Mod. Phys. D}, {\bf 18} (2009) 347
\bibitem{Fermi1} A.~A. Abdo, et al. {\it Astrophys. J.}, {\bf 706} (2009) 56
\bibitem{Fermi2} A.~A. Abdo, et al. {\it Astrophys. J.}, {\bf 701} (2009) 123
\bibitem{MW01} P. Meszaros and E. Waxman, {\it Phys. Rev. Lett.}, {\bf 87} (2001) 171102

\bibitem{LE93} A. Levinson and D. Eichler {\it Astrophys. J.}, {\bf 418} (1993) 386
\bibitem{GMA08} D. Giannios, P. Mimica and M.~A. Aloy, {\it Astron. Astrophys.}, {\bf 478} (2008) 747
\bibitem{Le09a} A. Levinson, {\it Astrophys. J.}, {\bf 705} (2009) 213

\bibitem{Vanetal04} M.~H. Van Putten, et al. {\it Phys. Rev. D}, {\bf 69} (2004) 044007
\bibitem{AD01} A. Atoyan and C.~D. Dermer {\it Phys. Rev. Lett.}, {\bf 87} (2001) 221102
\bibitem{LW01} A. Levinson and E. Waxman {\it Phys. Rev. Lett.}, {\bf 87} (2001) 171101
\bibitem{Dsetal02} C. Distefano, et al. {\it Astrophys. J.}, {\bf 575} (2002) 378
\bibitem{Chetal06} H.~R. Christiansen, M. Orellana and G.~E. Romero, {\it Phys. Rev. D}, {\bf 73} (2006) 063012
\bibitem{WB97} E. Waxman and J. Bachall {\it Phys. Rev. Lett.}, {\bf 78} (1997) 2292

\bibitem{Auger08} The Pierre Auger Collaboration, et al. 2008, Astropart. Phys., 29, 188
\bibitem{Kash08} T. Kashti and E. Waxman, {\it J. Cosmology Astropart. Phys.}, {\bf 5}, (2008) 6

\bibitem{BG99} E. Boldt and P. Gosh, {\it   Mon. Not. Roy. Astron. Soc.}, {\bf 307} (1999) 491
\bibitem{lev00} A. Levinson, {\it Phys. Rev. Lett.}, {\bf 85}, (2000) 912
\bibitem{RA09} F. Rieger and F. Aharonian, {\it Astron. Astrophys.}, {\bf 506} (2009) L41
\end{thebibliography}
\end{document}